\documentclass[3p]{elsarticle}

\usepackage{graphicx}
\usepackage[colorlinks=true]{hyperref}
\usepackage{amssymb}
\usepackage{siunitx}
\usepackage{upgreek} 
\usepackage{color} 
\definecolor{blue}{rgb}{0,0,0.8}

\definecolor{red}{rgb}{0.8,0,0}

\definecolor{green}{rgb}{0,0.4,0}

\definecolor{violet}{rgb}{0.8,0.2,0.8}
\definecolor{darkgreen}{RGB}{9, 90, 6}

\usepackage[normalem]{ulem}
\usepackage{multirow}
\usepackage{multicol}
\usepackage{nonfloat}

\definecolor{darkPurple}{RGB}{150,0,150}
\definecolor{lightBrown}{rgb}{0.7,0.7,0.5}

\usepackage{ifthen}
\newboolean{clean}
\setboolean{clean}{false}  

\ifthenelse{\boolean{clean}}
{
    \newcommand\commentOld[1]{\bigskip}
    
}
{
    \newcommand\commentOld[1]{{\color{lightBrown} #1}}

}

\usepackage{amsmath} 

\usepackage{lineno}
\usepackage{xspace}

\usepackage[symbol]{footmisc}

\biboptions{authoryear}

\journal{NeuroImage}

\makeatletter
\def\ps@pprintTitle{%
  \let\@oddhead\@empty
  \let\@evenhead\@empty
  \let\@oddfoot\@empty
  \let\@evenfoot\@oddfoot
}
\makeatother

\newcommand{\N}{\hat{\mathbf{n}}}
\newcommand{\U}{\hat{\mathbf{u}}}
\newcommand{\Q}{{\mathbf{q}}}
\newcommand{\B}{\mathbf{B}}
\renewcommand{\d}{\mathrm{d}}
\renewcommand{\v}{\mathbf{v}}
\newcommand{\la}{\left<}
\newcommand{\ra}{\right>}
\newcommand{\lb}{\left[}
\newcommand{\rb}{\right]}
\newcommand{\lp}{\left(}
\newcommand{\rp}{\right)}
\newcommand{\unit}[1]{\xspace\mathrm{#1}\xspace}

\begin{document}
\begin{frontmatter}
\title{Reproducibility of the Standard Model of diffusion in white matter on clinical MRI systems}

\author[1]{Santiago Coelho\corref{cor1}}
\ead{santiago.coelho@nyulangone.org}
\author[1]{Steven H. Baete}
\author[1]{Gregory Lemberskiy}
\author[1]{Benjamin Ades-Aaron}
\author[1]{Genevieve Barrol}
\author[1]{Jelle Veraart}
\author[1]{Dmitry S. Novikov}
\author[1]{Els Fieremans}
\address[1]{Bernard and Irene Schwartz Center for Biomedical Imaging, Department of Radiology, New York University School of Medicine, New York, NY, USA}
\cortext[cor1]{Corresponding author}


\begin{abstract}
Estimating intra- and extra-axonal microstructure parameters, such as volume fractions and diffusivities, has been one of the major efforts in brain microstructure imaging with MRI.  
The Standard Model (SM) of diffusion in white matter has unified various modeling approaches based on impermeable narrow cylinders embedded in locally anisotropic extra-axonal space. 
However, estimating the SM parameters from a set of conventional diffusion MRI (dMRI)  measurements is ill-conditioned.  Multidimensional dMRI helps resolve the estimation degeneracies, but there remains a need for clinically feasible acquisitions that yield robust parameter maps. Here we find optimal multidimensional protocols by minimizing the mean-squared error of machine learning-based SM parameter estimates for two 3T scanners with corresponding gradient strengths of $40$ and $80\,\unit{mT/m}$. We assess intra-scanner and inter-scanner repeatability for 15-minute optimal protocols by scanning 20 healthy volunteers twice on both scanners. 
The coefficients of variation all SM parameters except free water fraction are $\lesssim 10\%$ voxelwise and $1-4 \%$ for their region-averaged values. 
As the achieved SM reproducibility outcomes are similar to those of conventional diffusion tensor imaging, 
our results enable  robust in vivo mapping of white matter microstructure  in neuroscience research and in the clinic.  

\end{abstract}

\begin{keyword}
microstructure \sep Standard Model \sep diffusion \sep white matter \sep  experimental design \sep reproducibility
\end{keyword}

\end{frontmatter}


\begin{multicols}{2}

\section{Introduction}

\noindent
The promise of increased sensitivity and specificity in detecting brain microstructure changes is a major driving force for developing biophysical models in diffusion MRI (dMRI) \citep{JONES2010}. This imaging modality measures random displacements of water molecules within a voxel \citep{CALLAGHAN1991}, which are $\sim 10\,\mu$m in clinical experimental settings \citep{KISELEV2016}. Thus, dMRI images encode information about the tissue architecture restricting the diffusion of water molecules at a scale orders of magnitude below current MRI resolution \citep{NOVIKOV2019,ALEXANDER2019}, where disease processes originate. Brain microstructure mapping could provide biomarkers of pathological processes that would aid in early diagnosis \citep{ASSAF2008b}. This prompts the development of dMRI scan protocols and parameter estimation methods that are not only sensitive and specific, but also reproducible within clinically feasible scan time. 

\begin{figure*}[ht]
\centering
\includegraphics[width=\textwidth]{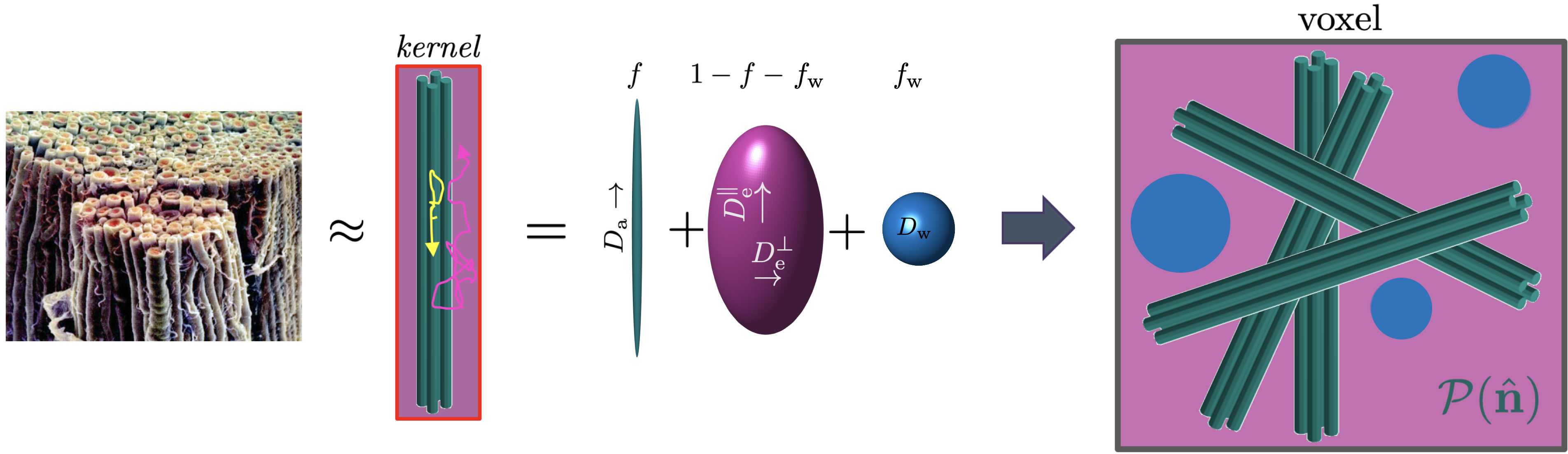}
\caption[caption FIG 1]{The Standard Model of diffusion in white matter. Shown is an elementary fiber fascicle (whose dMRI signal yields the so-called fiber response kernel), characterized by the compartment diffusivities and water fractions. A voxel is a collection of such fascicles oriented via an arbitrary fiber ODF ${\cal P}(\N).$
}\label{fig:SMkernel}
\end{figure*}

For water diffusion in brain white matter (WM), the overarching multiple  Gaussian compartment  framework is the so-called \textit{Standard Model} (SM), cf. Fig.~\ref{fig:SMkernel} and \citep{NOVIKOV2019} for a review. 
Briefly, axons (and possibly glial processes) are represented by impermeable zero-radius  cylinders (the so-called ``sticks") arranged in locally coherent fiber fascicles. The diffusion in the extra-axonal space of each fascicle is assumed to be Gaussian and described by an axially symmetric diffusion tensor. The third, optional tissue compartment is the cerebro-spinal fluid (CSF). Such multi-component fascicles are distributed in a voxel according to an arbitrary fiber orientation distribution function (ODF). All fascicles in a voxel are assumed to have the same compartment fractions and diffusivities, and differ from each other only by  orientation
(cf. Section \ref{s:Theory} for technical details).

The SM encompasses%
\footnote{The SM name originates from a tongue-in-cheek association with the Standard Model in the particle physics \citep{NOVIKOV2019}, as both encompass a fair bit of previous modeling effort from various groups. Here ``standard" refers to common assumptions among modeling approaches and does not imply ``exact". While particle physicists are on the lookout for physics beyond their SM, in dMRI such physics has been already found. At finite diffusion time, tissue compartments exhibit  residual non-Gaussian time-dependent diffusion. These effects, neglected in the SM, are below 10\% in clinical dMRI experiments. 
}
a number of WM models made of anisotropic Gaussian compartments with axons represented by sticks 
\citep{KROENKE2004,JESPERSEN2007,JESPERSEN2010,FIEREMANS2011,ZHANG2012,SOTIROPOULOS2012,JENSEN2016,JELESCU2015b,KADEN2016,REISERT2017,NOVIKOV2018,VERAART2017}. 
From the SM point of view, earlier models impose constraints either on compartment parameters or the functional form of the fiber ODF; such constraints improve robustness but may introduce biases into the estimation of remaining parameters (cf. recent reviews by \cite{JELESCU2017,NOVIKOV2019,ALEXANDER2019}). The constraints are typically employed when analyzing common dMRI acquisition protocols based on low-to-intermediate diffusion weightings with pulsed field gradient (PFG) measurements, since the  problem of recovering unconstrained SM parameters is ill-conditioned \citep{JELESCU2015b,NOVIKOV2018}.

Multidimensional diffusion MRI \citep{MITRA1995,WESTIN2016,TOPGAARD2017}
is a way to encode diffusion along more than one direction, probing the response to an ellipsoid encoded by a $3\times 3$  $\B$-tensor (cf. Figure \ref{fig:MDdMRI_TElandscape} and Section \ref{s:Theory} below).  
This adds complementary information to that accessible through conventional PFG, also known as linear tensor encoding (LTE) \citep{JESPERSEN2013,SZCZEPANKIEWICZ2016}. To resolve the degeneracy in SM parameter estimation, PFG/LTE has been combined with planar tensor encoding (PTE) at intermediate diffusion weightings \citep{COELHO2019,REISERT2019}. \cite{FIEREMANS2018} and \cite{DHITAL2018} analyzed the advantages of combining LTE with spherical tensor encoding (STE), and PTE, respectively, for SM parameter estimation. Similary, \cite{AFZALI2019} used numerical simulations to compare the estimation errors of the SM for a few discrete combinations of LTE, PTE and STE encodings. While all these studies show the value of multidimensional MRI for improving SM parameter estimation, some combinations of $\B$-tensor encodings  provide more precise and accurate estimation results than others.

Optimizing the parameter estimation is a complementary way to increase precision. This has made supervised machine learning (ML)-based approaches gain attention lately. Neural networks \citep{GOLKOV2016}, polynomial regression \citep{REISERT2017}, or random forest \citep{PALOMBO2020} have provided useful results in different dMRI applications. The ML approach has been applied to estimate SM parameters by \citep{REISERT2017}, and followed by more recent works  \citep{COELHO2021b,GYORI2021,DEALMEIDA2021b}. Irrespective of the implementation, all these works concluded that ML estimation alone is unable to resolve SM parameter degeneracies, and that a sufficiently rich acquisition protocol is needed. Furthermore, \cite{COELHO2021a} recently showed that as the signal-to-noise ratio (SNR) decreases, parameter estimates become increasingly influenced by the ML prior (the training set), and that an optimal acquisition minimizes such an undesired effect.

In this work we aim to use ML not only to enhance parameter estimation, but also to guide the experimental design. Acquisition optimization strategies are needed to reduce scan times and/or improve quality of SM parameter estimation. \cite{COELHO2019b} explored the space of isotropically distributed $\B$-tensors and selected the combination that maximized the precision of the cumulant tensor elements up to the second order in $\B$. Interestingly, this work did not impose measurements to be grouped in shells, but these emerged from the optimization. Subsequently, \cite{LAMPINEN2020} used the Cram{\'e}r-Rao bound (CRB) \citep{RAO1945,CRAMER1946} of the SM parameters to optimize the acquisition for an extended version of the SM that also accounts for intra-compartmental $T_2$ relaxation values \citep{VERAART2017}. While the CRB provides a lower bound on the variance for an unbiased estimator, this is not necessarily optimal for ML estimators which are usually biased at the expense of increased precision. Here we propose to use instead the (root) mean squared error (R)MSE as a quality metric for experiment design, as it enables trading off between bias and precision at each SNR. Hence, minimal MSE, rather than a proxy like the CRB, is a  target for the protocol design. Furthermore, the optimal protocol precision serves as a benchmark for the experimental precision in a reproducibility study, where other inevitable  factors, such as imaging artifacts and misregistration, can affect the net parameter variation.

Complementary to the search for the optimal acquisition, there has been an increasingly prescient need to evaluate reproducibility of dMRI, to enable its adoption in clinic and clinical research. \cite{GRECH2015} performed a multi-center reproducibility study of apparent diffusion coefficient (ADC) and diffusion tensor imaging (DTI) parameters with 1.5T and 3T MRI scanners, and found the relative error (coefficient of variation, or COV) of $<4\%$ for average values in different WM regions of interest (ROI). For diffusion kurtosis imaging (DKI) \citep{JENSEN2005}, \cite{HENRIQUES2021} recently proposed a regularized  estimator and studied its reproducibility on different publicly available datasets. Modeling the dMRI signal aims for more specific information than the above signal representations. \cite{ANDICA2020} assessed scan-rescan and inter-vendor reproducibility of neurite orientation dispersion and density imaging metrics \citep{ZHANG2012}, which is a constrained version of the SM. However, reproducibility studies for the unconstrained SM are so far lacking because they require non-conventional (beyond-LTE) diffusion data. 


The main outcomes of this work are (i) a framework that minimizes the estimation error of the SM parameters by coupling experiment design with ML-based parameter estimation, and (ii) its validation in a reproducibility study involving 20 healthy volunteers on two clinical scanners with different gradient strengths. 
In Section \ref{s:Theory} we introduce the SM adapted for multidimensional dMRI, and the ML-based parameter estimation. In Section \ref{s:Methods} we describe the experimental design, image acquisition, and processing. In Section \ref{s:Results} we report the resulting 15-minute optimal acquisition protocols for two 3T scanners with corresponding gradient strengths of $40$ and $80\unit{mT/m}$, Fig. \ref{fig:optimalProtocols}. 
We assess  the optimized protocols via numerical noise propagation and \textit{in vivo} experiments, Fig. \ref{fig:reprodMAPS}. Finally, in Section \ref{s:Discussion} we quantify the reproducibility of SM parameter estimates for 20 normal subjects in a scan-rescan on both scanners. Our optimized protocol achieves voxelwise $\rm{COV}\lesssim 10\%$ for all SM parameters except free water fraction; the COV for ROI-averaged values were $1-4\%$, Fig. \ref{fig:reprodSTATS}. These actual experimental values turn out to be in a good agreement with predictions based on the optimal MSE.

\section{Theory}\label{s:Theory}

\begin{figure*}[htbp]
\centering
\includegraphics[width=\textwidth]{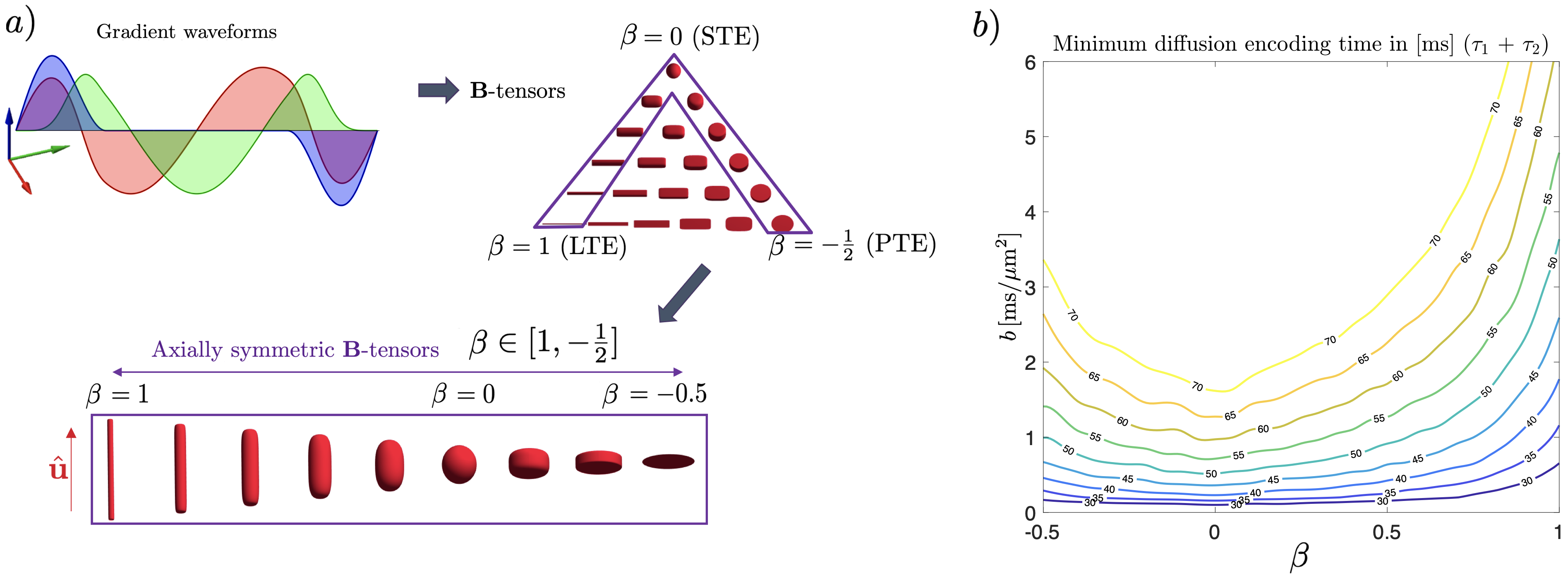}
\caption[caption FIG]{Elements in a multidimensional dMRI acquisition. a) shows a representation of an STE waveform $g(t)$ with the x (red), y (green), z (blue) axes. Superquadric glyphs representing $\B$-tensor shapes are arranged in a barycentric ternary diagram \citep{TOPGAARD2017}, according to their linear, planar, and spherical components. Axially symmetric $\B$-tensors lie on the edge of such diagram. b) Contour lines that show the maximum achievable b-value for a given $\B$-tensor shape and encoding time considering that $g_{\rm max}\leq 75\unit{mT/m}$ and slew rate $\leq 125\unit{mT/m/s}$. The latter is the sum of both encoding periods ($\tau_1 + \tau_2$), before and after the 180 degree radiofrequency pulse.}
\label{fig:MDdMRI_TElandscape}
\end{figure*}

\subsection{Multidimensional dMRI}\label{ss:MultidimdMRI}

\noindent
Multidimensional dMRI,  also known as $q$-space trajectory imaging (QTI)  \citep{ERIKSSON2013,WESTIN2016}, probes a trajectory $\Q(t)$, rather than a  point in the diffusion $q$-space, within a single measurement. In other words, QTI treats the dMRI signal as a {\it functional} of $\Q(t)$,  rather than a function of $\Q$, which results in   
the signal being sensitive to displacements along multiple dimensions simultaneously. 

For Gaussian diffusion, the picture gets simplified, as the cumulant series is truncated at the level of the second-order velocity cumulant $\la v_i(t)v_j(t')\ra = 2D_{ij}\delta(t-t')$ defining the diffusion tensor $D_{ij}$. Hence, the signal 
\begin{equation}\label{eq:S=eBD}
S[\Q(t)] = \la e^{i\int\! \d t\, \Q(t)\cdot\v(t)}\ra
= e^{-\sum_{ij} B_{ij} D_{ij}}
\end{equation}
gets reduced to a function of the $3\times 3$ symmetric tensor $\B=(B_{ij})$:
\begin{equation}\label{eq:Btensor}
\begin{aligned}
B_{ij} &= \int_0^{\text{TE}}\!\!q_i(t) \, q_j(t)\, \mathrm{d}t, \quad \text{with}\\
 q_i(t)&= \!\int_0^t \!\!g_i(t')\,\mathrm{d}t', \quad \text{and}\quad  b =  \sum_{i} B_{ii}\,,\\
\end{aligned}
\end{equation}
where the trace of $\B$ is the conventional  $b$-value\footnote{We use the ``microstructure units",  i.e., $\mu$m and ms, throughout the paper. In such units, the diffusion weighting, $b=1000\, \unit{s/ mm^2} = 1 \,\unit{ms/\mu m^2}$. 
In these units, free water diffusivity at body temperature $D_{\rm w}\approx 3 \unit{\mu m^2/ms}$.}, 
and $\mathbf{g}(t) = \mathrm{d}\Q/\mathrm{d}t$ is the diffusion gradient waveform used to generate the trajectory.
Hence, for a medium (\textit{e.g.}, a voxel) comprised of 
multiple non-exchanging Gaussian-diffusion compartments (in general, anisotropic, such as in the SM), the $\B$-tensor fully parametrizes the measurement.

The number of nonzero eigenvalues in $\B$ reflects how many dimensions of the anisotropic Gaussian diffusion are being probed simultaneously. In this work, we focus on axially symmetric $\B$:
\begin{equation}\label{eq:AxSymB}
B_{ij}(b,\beta,\U)=b\left( \beta \, u_i u_j + \frac{1-\beta}{3} \delta_{ij} \right)
\end{equation}
that are parametrized by  the overall scale (the $b$-value), the unit vector $\U$ along the symmetry axis, and the shape parameter $\beta$ (Fig. \ref{fig:MDdMRI_TElandscape}a). 
Compared to conventional dMRI, the extra degree of freedom $\beta$ represents the $\B$-tensor shape, \textit{e.g.}, $\beta=1$ for linear encoding (a single nonzero eigenvalue), $\beta=0$ for spherical encoding (isotropic $\B$-tensor),  and $\beta=-\tfrac12$ for planar encoding (two nonzero eigenvalues). 
Here $\delta_{ij}$ is the Kronecker symbol (the unit matrix). 


The encoding time required for the gradient waveforms corresponding to a given $\B$-tensor depends on its shape and b-value. Figure \ref{fig:MDdMRI_TElandscape}b shows the minimum diffusion encoding times for $\{b,\beta\}$ combinations given specific hardware constraints (more details in Section \ref{ss:OptimalExpDesign}). Isotropic weighting and large $b$-values demand more encoding time if gradient hardware constraints are kept constant.


\subsection{SM for multidimensional dMRI}
\label{ss:StandardModel}

\noindent
Multiple  approaches \citep{KROENKE2004,JESPERSEN2007,JESPERSEN2010,FIEREMANS2011,ZHANG2012,SOTIROPOULOS2012,JENSEN2016,JELESCU2015b,KADEN2016,REISERT2017,NOVIKOV2018,VERAART2017}
to model the physics of water diffusion in WM had relied on similar assumptions. This led to the unifying framework dubbed \textit{Standard Model} (SM) of diffusion in WM
as formulated in \citep{REISERT2017,NOVIKOV2018,NOVIKOV2019}. 

Consider an elementary fiber segment or fiber fascicle, which is a local bundle of aligned sticks with the extra-neurite space surrounding them. 
The  signal from such fascicle oriented along the unit vector $\N$, has contributions from two axially symmetric non-exchanging Gaussian compartments aligned along $\N$, Fig.~\ref{fig:SMkernel}:
\begin{itemize}
    \item {\it Stick compartment}, with signal fraction $f$, representing axons and possibly other elongated cells such as glial processes. Sticks are zero-radius cylinders, where diffusion occurs only along the cylinder axis with diffusivity $D_\text{a}$, such that the diffusion tensor is $D_\text{a} \, n_i n_j$. 
    \item {\it Zeppelin compartment}  reflecting hindered diffusion in the extra-axonal space. Its diffusion tensor eigenvalues are the parallel and perpendicular diffusivities $D_\text{e}^\parallel$ and $D_\text{e}^\perp$,  such that the tensor is $\Delta_\text{e} \,n_i n_j + D_\text{e}^\perp\delta_{ij}$, where $\Delta_\text{e} = D_\text{e}^\parallel-D_\text{e}^\perp$.
    \item {\it Free water compartment}  with signal fraction $f_\text{w}$ and fixed isotropic diffusivity $D_\text{w}=3\,\unit{\mu m^2/ms}$ is optionally added to account for CSF partial volume contributions. We will include it in our analysis. 
\end{itemize}

Applying Eqs.~(\ref{eq:S=eBD}) and (\ref{eq:AxSymB}) to each compartment above, and using 
\begin{align}\nonumber
\sum_{ij}
\lp \lambda n_i n_j + \mu\delta_{ij}\rp 
\lp\beta u_i u_j + \frac{1-\beta}3\delta_{ij}\rp
\\ \nonumber
=
\beta\lambda \lb\xi^2-\frac13\rb + \frac{\lambda}3 + \mu\,, 
\end{align}
where $\xi = \N \cdot \U$ is the cosine of the angle between the symmetry axes of the kernel and of the $\B$-tensor,
we obtain a fascicle's response function (or a response kernel) to the measurement encoded by the $\B$-tensor: 
\begin{equation}\label{eq:KernelTensor}
\begin{aligned}
\mathcal{K}(b,\beta,\xi) 
    = f \, &\exp\bigl[- bD_{\text{a}} \big(   \beta (\xi^2-\tfrac13) + \tfrac13 \big) \bigr] 
    	\\
    	+ (1-f-f_\text{w})  &\exp\bigl[- bD_\text{e}^\perp - b\Delta_\text{e} \big(   \beta (\xi^2-\tfrac13) + \tfrac13 \big) \bigr] 
    	\\
    	+ f_\text{w}  &\exp\bigl[- b D_{\text{w}} \bigr] .
\end{aligned}
\end{equation}

Voxels contain not one but a collection of fiber segments whose orientation is given by a probability distribution on the sphere $\mathcal{P}(\N)$, dubbed fiber orientation distribution function (ODF). Thus, the SM signal becomes the convolution of the kernel and the ODF on the unit sphere%
\footnote{Technically, the convolution is defined on the rotation group SO(3), equivalent to the 3-dimensional unit sphere $\mathbb{S}^3$.  
However, due to the fiber fascicle's axial symmetry, the SO(2) rotation around its axis can be factored out, and the convolution becomes over the factor group SO(3)/SO(2) equivalent to the 2-dimensional unit sphere $\mathbb{S}^2$, $||\N||=1$ \citep{HEALY1998}. 
Henceforth, as in  \citep{NOVIKOV2018}, we normalize the measure on $\mathbb{S}^2$, 
$\d \N \equiv \sin\theta\,  \d\theta\, \d\phi / 4\pi$, 
such that $\int\! \d \N = 1$.}:
\begin{equation}\label{eq:IntegratedKernelTensor}
S(\B)\! =\! 
         s_0 \!\int_{\mathbb{S}^2}\! \mathrm{d}\N\,  
         \mathcal{P}(\N)\, 
        \mathcal{K}(b,\beta,\N \cdot \U) \,,
\end{equation}
where $s_0 \equiv S(\B)|_{\B=0}$ is the non-weighted  signal, and
$\mathcal{P}(\N)$ is normalized to the unit probability,  $\int_{\mathbb{S}^2}\mathcal{P}(\N)\,\mathrm{d}\N=1$.

The SM ODF is represented via a spherical harmonic (SH) decomposition:  
\begin{equation} \label{ODF_SH}
    \mathcal{P}(\N) \approx 1+\sum_{\ell=2,4, \ldots}^{\ell_{\max }} \sum_{m=-\ell}^{\ell} p_{\ell m}\, Y_{\ell m}(\N)\,,
\end{equation}
where $Y_{\ell m}(\N)$ are the SH basis functions conventionally normalized to 
\[
4\pi\int\!\d\N\, Y^*_{\ell m}(\N)Y_{\ell' m'}(\N) = \delta_{\ell\ell'}\delta_{mm'}\,, 
\]
$p_{\ell m}$ are the SH coefficients 
(only even $\ell$ are nonzero due to the time-reversal symmetry of the Brownian motion and $p_{00}=\sqrt{4\pi}$), and $\ell_{\max }$ is the maximum order in the expansion (typically 4--8, depending on the SNR and the maximal $b$).

The ODF form (\ref{ODF_SH}) is chosen due to a number of reasons. First, the SH basis is standard for functions on a sphere, and it does not give preference to any particular functional form of the ODF (since there are currently no justifiable empirical ODF models). Second, this basis realizes the angular-radial connection in the $q$-space \citep{NOVIKOV2018} (the successive terms in the Taylor expansion of Eq.~(\ref{eq:IntegratedKernelTensor}) in the powers of $q_{i_1}\dots q_{i_\ell} \sim b^{\ell/2}$ correspond to the sensitivity to SH up to order $\ell$).  Third, the convolution on a sphere becomes a product in the SH basis, as discussed below.   
As a result, the free parameters of the SM are the compartmental diffusivities and water fractions of the kernel, and the fiber ODF coefficients $p_{\ell m}$. 

When referring to the SM, unless specified otherwise, it is implied that kernel and ODF parameters are estimated directly from the data without constraints on parameters \citep{NOVIKOV2019}. Kernel parameters provide important tissue microstructural information, and have shown potential clinical relevance as they are sensitive to specific disease processes such as demyelination \citep{FIEREMANS2012,JELESCU2016} ($D_\text{e}^\perp$), axonal loss \citep{FIEREMANS2012} ($f$) or beading \citep{LEE2020} ($D_\text{a}$). Additionally, we anticipate that ODF parameters ($p_{\ell m}$) can be used for a more accurate tractography since the kernel is estimated at each voxel locally, rather than being averaged over  white matter tracts as in model-free deconvolution methods \citep{TOURNIER2004}.

\subsection{ODF-kernel factorization and rotational invariants}\label{ss:RotInvs}

\noindent
The Fourier transform diagonalizes the convolution operation. In other words, it provides a basis where convolutions become products. 
The Fourier basis on a sphere is the SH basis (\ref{ODF_SH}). In this basis, the convolution in Eq.~(\ref{eq:IntegratedKernelTensor}) becomes a product: 
%
%
\begin{equation}\label{eq:lmFactorization}
    S_{\ell m} (b,\beta) = s_0\, p_{\ell m}\, \mathcal{K}_\ell(b,\beta)\,.
\end{equation}
Here 
\begin{equation}
    \mathcal{K}_\ell(b,\beta) = \int_0^1 d\xi \,  \mathcal{K}(b,\beta,\xi) P_\ell (\xi)
\end{equation}
are the projections of the kernel onto the Legendre polynomials  $P_\ell(\xi)$ (proportional to $m=0$ SH), 
such that 
\begin{equation}\label{KlPl}
    \mathcal{K}(b,\beta,\N\cdot\U)= \sum_{l=0,2,\dots}
    (2l+1)\,\mathcal{K}_\ell(b,\beta)
    P_\ell(\N\cdot\U)\,.
\end{equation}
Indeed, substituting Eqs.~(\ref{ODF_SH}) and (\ref{KlPl}) into Eq.~(\ref{eq:IntegratedKernelTensor}), and using the SH addition theorem
\[
P_\ell(\N\cdot\U) = {4\pi\over 2l+1}
\sum_{m=-\ell}^\ell 
Y^*_{\ell m}(\N) Y_{\ell m}(\U) \,, 
\]
one gets the signal SH coefficients (\ref{eq:lmFactorization}). 
Since the fascicle is axially symmetric and is probed by an axially symmetric $\B$-tensor,  the $m\neq0$ SH coefficients 
$\mathcal{K}_{\ell m}$ of the kernel vanish. 


To remove the dependence on the choice of the physical basis in three-dimensional space (via $m=-\ell ... \ell$),  the  rotational invariants of the signal and ODF  are employed: 
\begin{equation}\label{eq:RotInvs}
\begin{aligned}
S^2_\ell(b,\beta) &= \frac{1}{4\pi (2\ell + 1)}\sum_{m=-\ell}^{\ell} |S_{\ell m}(b,\beta)|^2, \\
p_\ell^2 &= \frac{1}{4\pi (2\ell + 1)}\sum_{m=-\ell}^{\ell} |p_{\ell m}|^2.
\end{aligned}
\end{equation}
The above normalization is chosen such that: i) $p_0=1$, since the integral of the ODF over the sphere equals 1; ii) the remaining ODF invariants characterizing anisotropy satisfy $0 \leq p_\ell \leq 1$. It follows that for $\ell \geq 2$ an isotropic ODF has $p_{\ell} = 0$ while a delta-function on a sphere (a perfectly aligned fiber tract) has all $p_{\ell} = 1$. Hence, $p_\ell$ are the ``partial" ODF anisotropy metrics at each degree $\ell$. 

From Eqs.~(\ref{eq:lmFactorization}) and (\ref{eq:RotInvs}), one can relate the signal rotational invariants to the kernel parameters:
\begin{equation}
    S_{\ell} (b,\beta) = s_0\,  p_{\ell} \,\mathcal{K}_\ell(b,\beta),\quad \ell=0,2,... \, .
\end{equation}
This allows separating the parameter estimation in two steps $\{S(b,\beta,\U)\} \rightarrow \{S_{\ell}(b,\beta)\} \rightarrow \text{kernel}$, without loss of information and having to estimate only a few $p_\ell$ from the ODF. 
The above treatment generalizes earlier factorization approach \citep{REISERT2017,NOVIKOV2018} from LTE to arbitrary axially-symmetric $\B$-tensors.

\subsection{Supervised machine learning regression}\label{ss:MLRegression}

\noindent
Unlike conventional parameter estimation approaches which rely on an analytical forward model, \textit{e.g.} maximum likelihood, data-driven ML regressions \textit{learn} the mapping from noisy measurements to model parameters. This is done by applying a sufficiently flexible regression to training data generated with the forward model of interest, which generally contains noise to mimic realistic scenarios. In the SM context, this approach was pioneered by \cite{REISERT2017}. Interestingly, \cite{COELHO2021a} showed that the learned mapping becomes smoother with increased levels of noise, effectively removing high order features that would be obtained in the case of noise-free mapping. Thus, for typical SNR values found in clinical dMRI experiments, the optimal regression, \textit{i.e.} minimizing MSE of the training data, can be achieved already by a cubic polynomial (suggested by \cite{REISERT2017} for LTE), as it  captures all relevant degrees of freedom in the data represented by the set of $S_\ell$.

A major advantage of polynomial regression over neural networks is that it is a linear optimization problem and the training can be computed much faster without the risk of local minima. \cite{COELHO2021a}  provided fast analytical equations to compute not only optimal regression coefficients but also the MSE over a distribution of values for all model parameters with a given acquisition protocol and noise level. 
This makes MSE a good metric for comparing the performances of different protocols and thus, a better objective function than the CRB for optimal experimental design.


\section{Methods}\label{s:Methods}
\subsection{Optimal experimental design}\label{ss:OptimalExpDesign}

\noindent
To select the experimental design that exploits our parameter estimation the most, we look for the set of measurements that minimize the RMSE in the parameters estimates:
\begin{equation}\label{Eq_MSE}
    \text{RMSE} = \sqrt{(\text{bias})^2 + \text{variance}}\,,
\end{equation}
where $\sqrt{\text{variance}}$ typically scales with the noise level. By minimizing Eq.~(\ref{Eq_MSE}) we are simultaneously aiming for increased accuracy and precision. 


The metric for quantifying the goodness of a protocol was: 
$\text{RMSE}_\text{obj} = \text{RMSE}_f + \tfrac13 \text{RMSE}_{D_\text{a}} + \tfrac13 \text{RMSE}_{D_\text{e}^\|} + \tfrac13 \text{RMSE}_{D_\text{e}^\perp} + \text{RMSE}_{f_\text{w}}$, where each parameter is normalized by its range. This assured an even sensitivity to all parameters was kept. Individual MSE values were computed analytically as in \cite{COELHO2021a}, based on the measured shells, the noise level, and the moments of the training data distribution. 
This accurately captures how experimental design and SNR affect our parameter estimates over a distribution of values.

We assume that the acquisition time is fixed and defined by the user. In this work we focus on 15-minute scan times. Thus, the optimization framework had to find the best way to fill the available time without violating hardware limitations such as maximum gradient amplitude and slew rate. Shells of uniformly distributed directions were assumed to be part of the optimal acquisition since it has been shown that measurements grouped into shells increase precision \citep{COELHO2019b}. For each shell, the optimization framework selected: diffusion weighting $b$, the $\B$-tensor shape parameter $\beta$, number of directions, and TE.

Using the framework proposed by \cite{SJOLUND2015,SZCZEPANKIEWICZ2019} we generated a library of 
minimum encoding times as a function of the $\{b,\beta\}$ combination, see Fig. \ref{fig:MDdMRI_TElandscape}b. These were used to compute how much encoding time was needed for each specific $\B$-tensor on each scanner. 
The optimization handled the trade-off between adding many shorter high-SNR measurements or fewer longer low-SNR ones. The TE was constrained to be equal for all shells to factor out $T_2$ dependence in the analysis. Thus, the overall TE, and the SNR of the dataset, was determined by the longest diffusion waveform.

We included conventional DKI shells (LTE, $b=0-1-2 \unit{ms/\mu m^2}$, \citep{SOTIROPOULOS2013,CASEY2018,ALFAROALMAGRO2018}) as fixed into our protocols. This enables future comparisons against standard DKI-derived maps and increases the flexibility of the data analysis. 
The maximal $b$-value that the optimizer could explore was set to $b_{\text{max}}=10 \unit{ms/\mu m^2}$ and $b_{\text{max}}=8 \unit{ms/\mu m^2}$ to accommodate our two scanners described below. Stochastic optimization \citep{ZELINKA2004} was used to navigate the high-dimensional and non-convex protocol landscape.

\subsection{\textit{In vivo} dMRI experiments}

\noindent
Twenty normal volunteers (23-66 years old, 10 males - 10 females) underwent brain diffusion MRI on Siemens Magnetom Prisma and Skyra 3T systems ($80\unit{mT/m}$ and $40\unit{mT/m}$ gradient systems, respectively), using a 20-channel head coil. The local Institutional Review Board approved the study and informed consent was obtained and documented from all participants. Maxwell-compensated asymmetric waveforms \citep{SZCZEPANKIEWICZ2019} were employed in all acquisition protocols using an in-house diffusion sequence with EPI readout \citep{VERAART2019} and with a single TE. Isotropically distributed directions were used at different combinations of diffusion weightings and encodings (see Fig. \ref{fig:optimalProtocols} for a representation of both protocols). Both protocols took 15 minutes to acquire for each repetition.

On the Skyra scanner, scan(1)-rescan(2) of this protocol was acquired with TR/TE=6700/127ms (SKYRA$_\text{127}^\text{(1)}$ and SKYRA$_\text{127}^\text{(2)}$), while on the Prisma scanner, scan(1)-rescan(2) was acquired with TR/TE=5300/92ms (PRISMA$_\text{92}^\text{(1)}$ and PRISMA$_\text{92}^\text{(2)}$) in addition to one scan(3) matching the Skyra protocol with TR/TE=6700/127ms (PRISMA$_\text{127}^\text{(1)}$). Imaging parameters: resolution: $2.0\unit{mm}$ isotropic, in-plane FOV: 220mm, GRAPPA and SMS acceleration factors: 2, $\text{PF}=6/8$. Subjects were taken out of scanner and repositioned between scans.

\subsection{Image pre-processing}

\noindent
Magnitude and phase data were reconstructed using projection onto convex sets (POCS) reconstruction \citep{HAACKE1991}. Then, a phase estimation and unwinding step preceded the denoising of the complex images \citep{LEMBERSKIY2019}. Denoising was done using the Marchenko-Pastur principal component analysis method \citep{VERAART2016}. By preserving only the significant principal components in the signal, this method reduces the noise with minimal smoothing. An advantage of denoising before taking the magnitude of the data is that Rician bias is reduced significantly. 

Data was subsequently processed with the DESIGNER pipeline \citep{ADESARON2018}. Denoised images were corrected for Gibbs ringing artifacts
\citep{KELLNER2015,TOURNIER2019}, based on re-sampling the image using local sub-voxel shifts. These images were rigidly aligned and then corrected for eddy current distortions and subject motion simultaneously \citep{SMITH2004}. A $b=0$ image with reverse phase encoding was included for correction of EPI-induced distortions \citep{ANDERSSON2003}.

\subsection{Parameter estimation}

\noindent
We used the regression method described in Section \ref{ss:MLRegression} where the SM parameters are estimated from the rotational invariants, 
employing the cubic polynomial regression of \cite{REISERT2017}. Compared to fitting directly the diffusion-weighted signals, this has the advantage of reducing the dimensionality of the problem without loss of generality, since rotational invariants are model-free and capture all the SM microstructural information in a few variables. 
For SNR ranges accessible with diffusion MRI there is no gain for using a more complex regression such as neural networks (\textit{cf.} \cite{COELHO2021a}). Furthermore, polynomial regressions are convex problems which are faster to solve than other types of machine learning regressions. The rotational invariants were estimated from the diffusion-weighted images using linear least squares since these had minimal bias after denoising complex images. All codes for SM parameter estimation were implemented in MATLAB (R2021a, MathWorks, Natick, Massachusetts). These are publicly available as part of the SMI (standard model imaging) toolbox at \href{https://github.com/NYU-DiffusionMRI/SMI}{https://github.com/NYU-DiffusionMRI/SMI}. DKI parameters were estimated with weighted linear least-squares \citep{VERAART2013} on the DKI shells ($\sim 40\%$ of the acquired data).

For each kernel parameter $x$, the coefficients of variation $\mbox{COV}=\sigma/\mu$ were computed from the two repetitions $x_{1,2}$ we had for each protocol, where the mean and standard deviation were estimated as: $\hat{\mu}=\tfrac12(x_1+x_2),\,\hat{\sigma} = \tfrac{\sqrt{\pi}}{2}  |x_1 - x_2|$. 
The, COV values were averaged over either voxels or over ROIs.

\section{Results}\label{s:Results}

\subsection{Protocol optimization}\label{ss:OptimalProtocols}

\begin{figure*}[htbp]
\centering
\includegraphics[width=\textwidth]{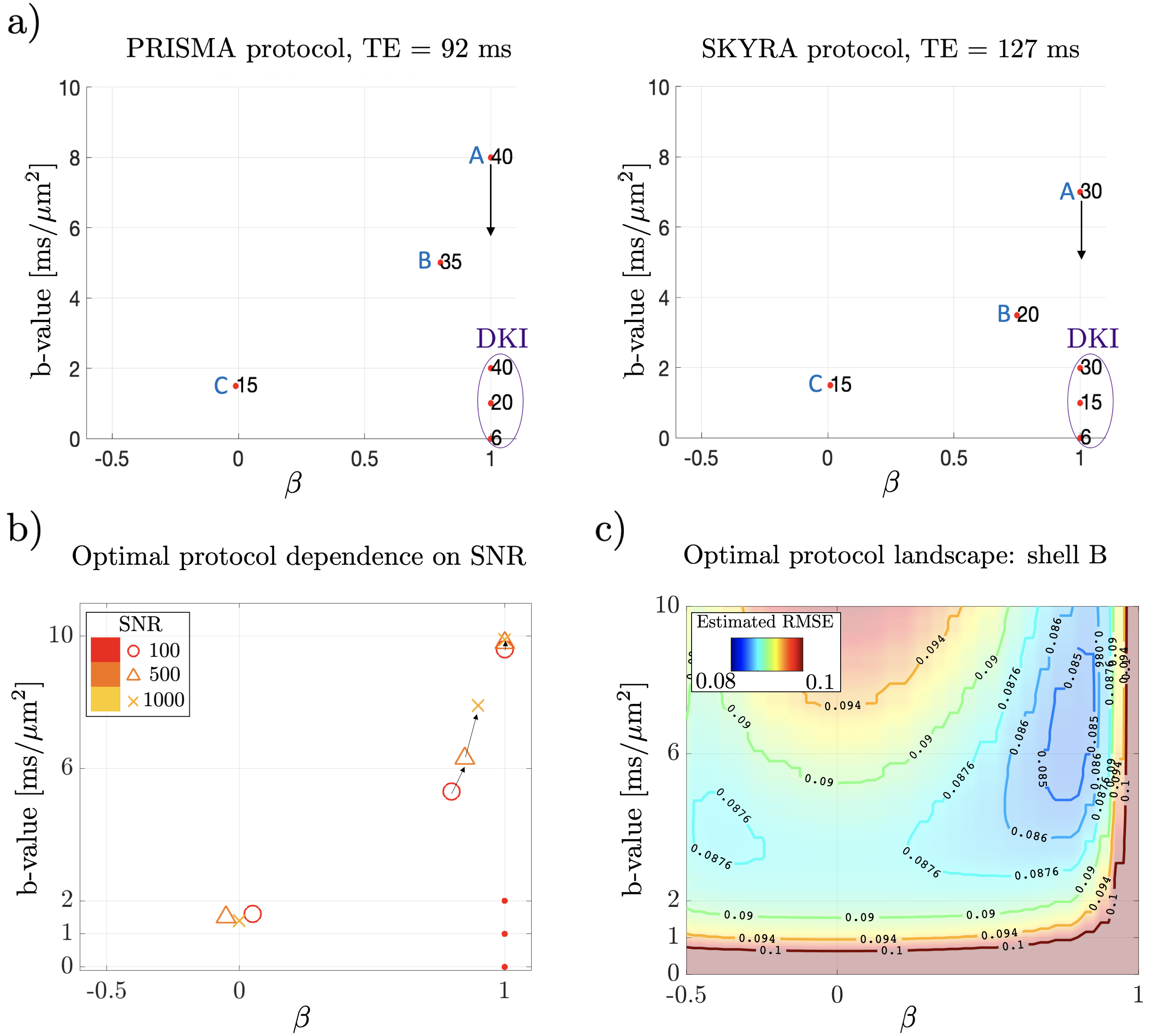}
\caption[caption FIG]{
a) Results from the optimization search for the two scanners. Each red dot represents a shell with its unique combination of b-value, $\B$-tensor shape, and number of measurements. Although these optimizations considered different hardware constraints, they look qualitatively similar. In both cases DKI shells were fixed and the optimization selected the remaining ones (A, B, and C). Black arrows indicate the decreased b-value for the high-b LTE shell to avoid gradient duty cycle issues. 
b) Optimal protocols for different SNR levels. The positions of shells A and B move towards higher diffusion weightings as the SNR increases. 
c) Landscape of the RMSE objective function (average over parameters normalized with the parameter range) for the position of the shell B, after fixing DKI shells and shell A.
Shell C is not present for simplicity, to illustrate the RMSE improvement due to going beyond the linear encoding.
}
\label{fig:optimalProtocols}
\end{figure*}

\noindent
Figure \ref{fig:optimalProtocols}a shows a representation of the optimized protocols, with b-values, b-shapes and number of directions for each shell. The proposed optimal acquisitions were similar for both scanners ($40$ and $80\unit{mT/m}$ maximum gradient strength).  After fixing DKI shells into the optimization problem (LTE, $b=0,1,2\unit{ms/\mu m^2}$), we see that {\it three complementary shells arise:} 
\begin{itemize}
    \item 
    (A) a high-$b$ LTE shell ($b\geq5 \unit{ms/\mu m^2},\, \beta\! =\!1$);  
    \item 
    (B) an intermediate/high-$b$ highly anisotropic $\B$-shape shell ($3.5\leq b\leq5 \unit{ms/\mu m^2}$, $0.75 \!\leq\! \beta\! \leq\! 0.8$);  
    \item 
    (C) an intermediate-$b$ STE shell ($b=1.5 \unit{ms/\mu m^2},\, \beta\! =\!0$). 
\end{itemize}
These locations in the acquisition space provide \textit{functionally independent} forms for the rotational invariants while maintaining a sufficiently high SNR. Figure \ref{fig:optimalProtocols}b shows how the positions of the non-DKI shells vary as SNR increases.

Non-LTE data contributes significantly to parameter precision. As an example, Fig. \ref{fig:optimalProtocols}c shows the normalized and averaged RMSE of all model parameters after fixing the DKI shells and high-b LTE shell. Here, we move the position of the shell B in the full $\{b,\beta\}$ acquisition space. We remove shell C, and keep TE and number of directions fixed for simplicity. It can be appreciated that small modifications will not affect the protocol significantly since the RMSE landscape is very flat in their neighborhood. However, the sharp transition of the averaged RMSE as shell B moves away from LTE in Fig. \ref{fig:optimalProtocols}c makes evident that after adding high-$b$ LTE data to the DKI shells, non-LTE shells are the most informative.


\subsection{Volunteer experiments}\label{ss:VolunteerResults}

\begin{figure*}[htbp]
\centering
\includegraphics[width=\textwidth]{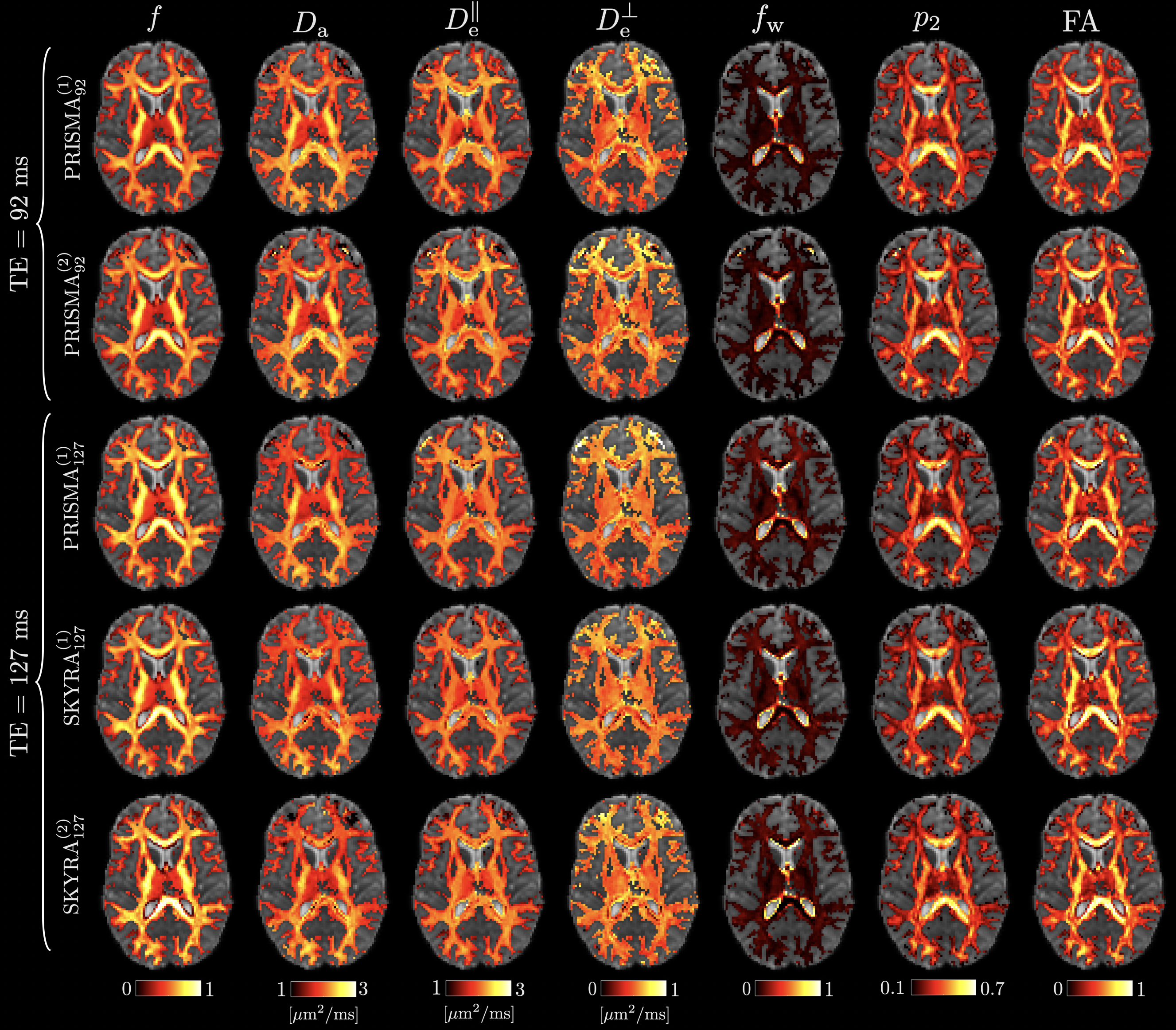}
\caption[caption FIG]{Columns show different SM parameter and fractional anisotropy (FA) maps for a 25 year old female control subject. Color maps of SM WM 
parameters are plotted on top of a $T_2$-weighted image. Each row contains a different repetition, labeled as SCANNER$_\text{TE}^\text{repetition}$. The first two rows were computed at a TE=92ms (protocol optimized for the PRISMA scanner) while the bottom three rows were computed at a TE=127ms (protocol optimized for the SKYRA scanner).}
\label{fig:reprodMAPS}
\end{figure*}

\begin{figure*}[htbp]
\centering
\includegraphics[width=\textwidth]{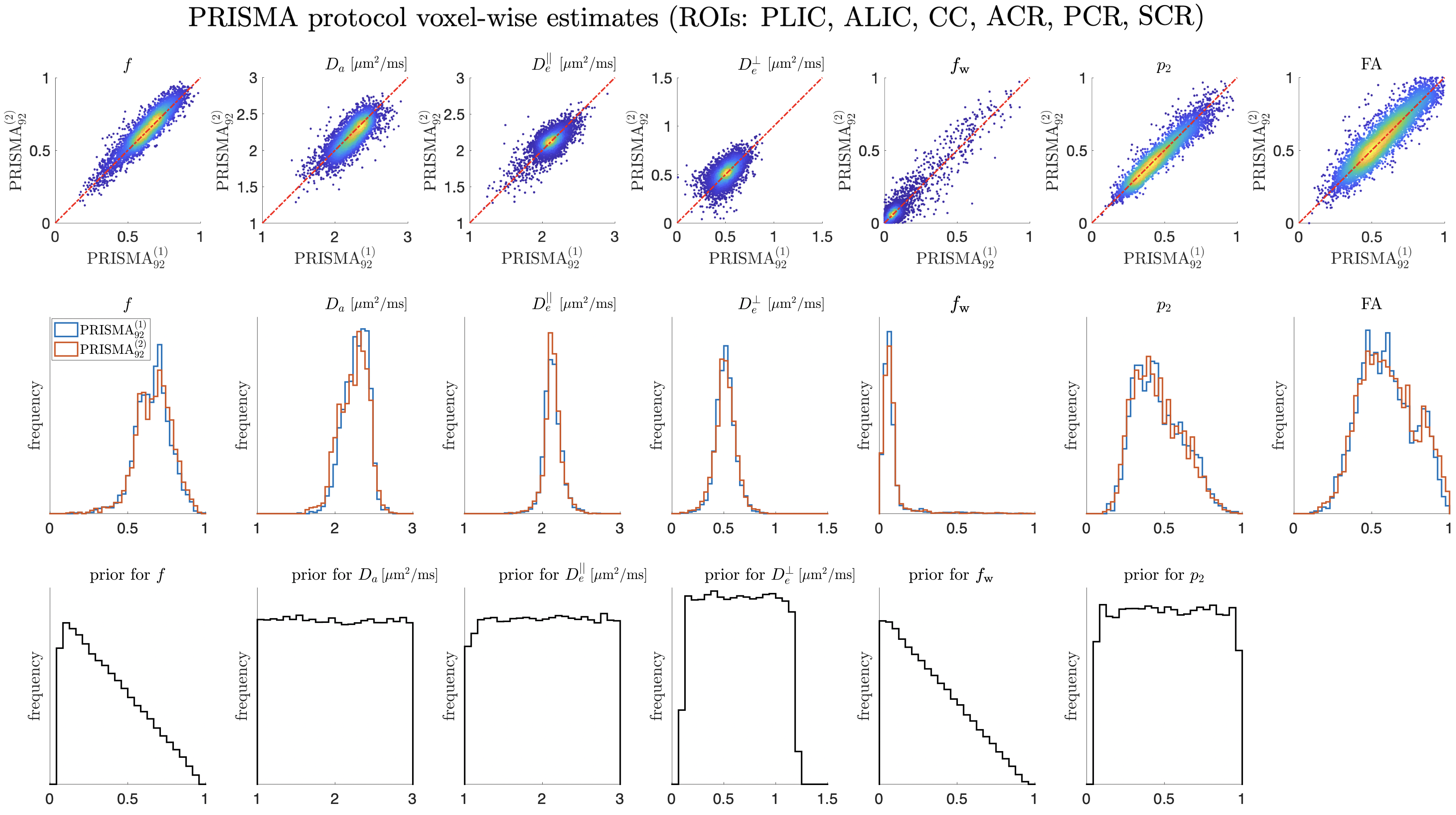}
\caption[caption FIG]{Voxel-wise reproducibility of Standard Model parameters and FA for the PRISMA scan-rescan. The bottom row shows the prior distributions used in the machine learning regression. Uniform distributions were chosen for all diffusivities and $p_2$. Since $0\leq f+ f_\text{w} \leq 1$, $0\leq f\leq 1$, and $0\leq f_\text{w}\leq 1$, the less informative prior is assuming the sum of water fractions being uniformly distributed between 0 and 1. This makes their individual distributions triangular.}
\label{fig:ROIvalues_wprior}
\end{figure*}

\begin{figure*}[htbp]
\centering
\includegraphics[scale=.20]{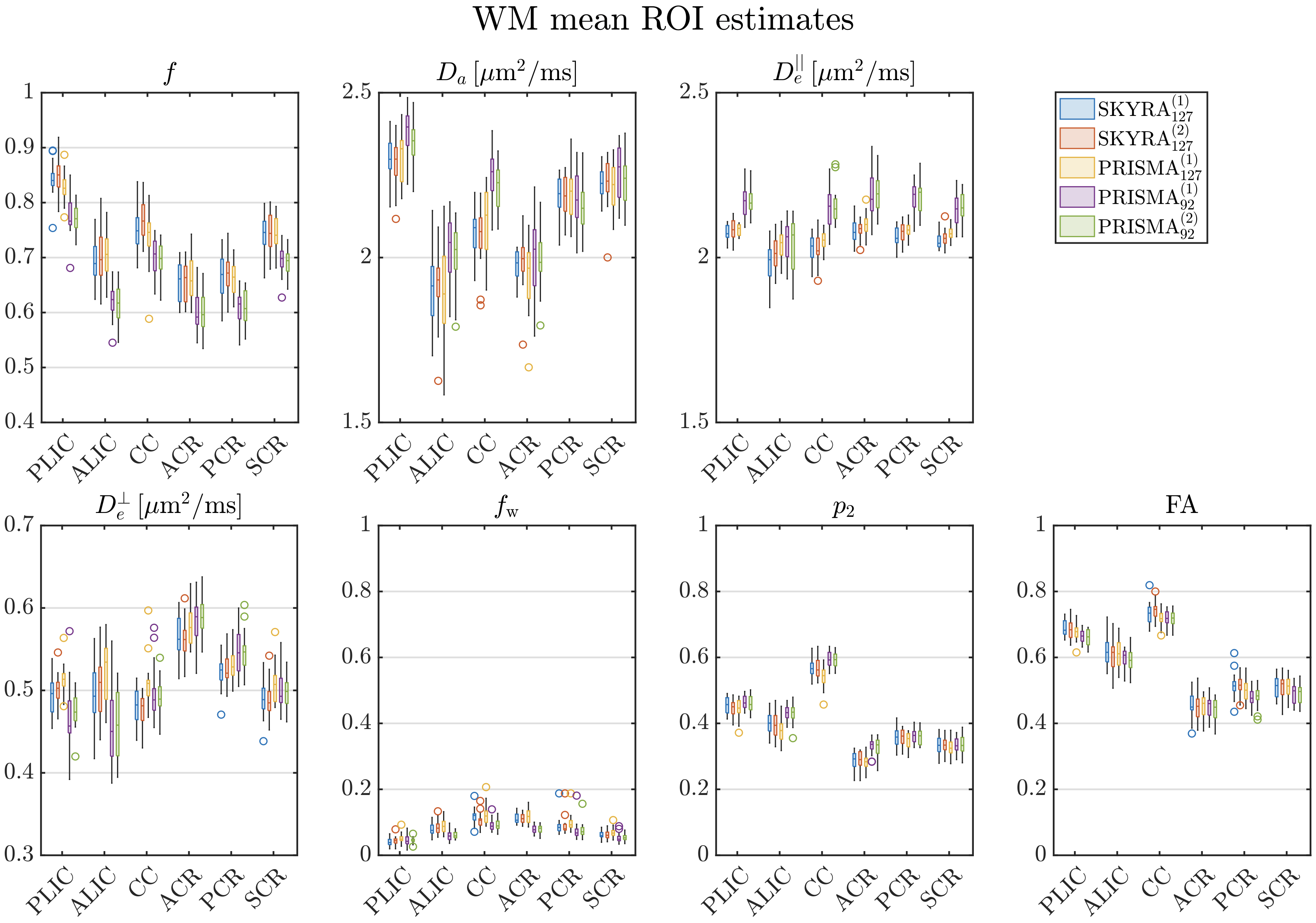}
\caption[caption FIG]{Box plots showing mean values for all subjects for all SM parameters for different WM ROI.Biological variability of the different SM parameters is observed across the WM. Water fractions and dispersion ($p_2$) are, as expected, different between protocols of different TE due to different $T_2$-weighting. However, we also observe a small decrease of the parallel diffusivities at larger TE values.}
\label{fig:ROIvalues_boxplots}
\end{figure*}

\noindent
Representative WM parametric maps are shown in Fig. \ref{fig:reprodMAPS} for a 25 year old female subject. Maps are visually reproducible for both TEs, even when comparing PRISMA vs SKYRA. 
Reproducibility on both scanners and between scanners showed similar results but the best COVs were seen on the Prisma due to better performing gradients providing a higher SNR due to a shorter TE (see representative histograms of WM voxels in Fig. \ref{fig:ROIvalues_wprior} and ROI average values in Fig. \ref{fig:ROIvalues_boxplots}). Here, voxelwise COVs were between $5-10\%$ for $f$, $D_\text{a}$, $D_\text{e}^{\perp}$, and $p_2$. $D_\text{e}^{||}$ showed the smallest COV ($\sim 3\%$), possibly due to a combination of having small variability and being the hardest parameter to estimate, making it the most influenced by the training data \citep{COELHO2021a}. Note that $f_\text{w}$ is not only a noisy parameter but it also has small values, leading to very high relative variations ($\sim 40\%$). 

\begin{figure*}[htbp]
\centering
\includegraphics[scale=0.45]{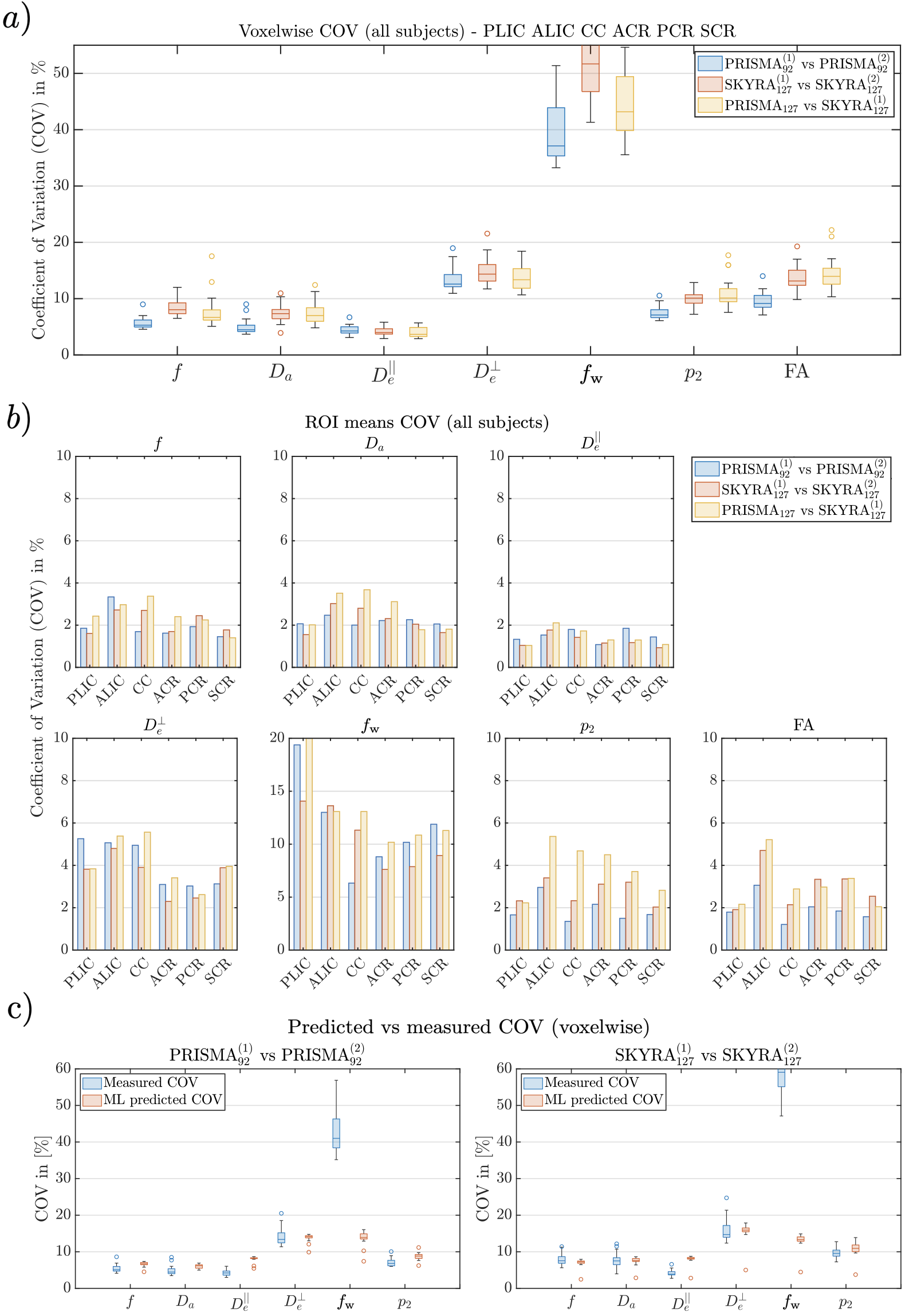}
\caption[caption FIG]{Coefficients of variation (COV) for all SM parameters and FA (computed from the DKI shells). a) shows measured COVs for voxel reproducibility. b) Shows COVs for the means of the main WM ROIs. Note that due to its low value, free water fraction has a large COV since this is a relative error metric. c) Comparison of the measured voxelwise COVs against the predicted ones through a noise propagation experiment.}
\label{fig:reprodSTATS}
\end{figure*}

The COV of SM parameters are comparable to the ones from FA estimated from the DKI subset. This is encouraging since FA is known to be reproducible due to being  
a much simpler parameter to measure and estimate (intra-scanner voxelwise COV $\simeq 7-13\%$, ROI COV $\simeq 1-4\%$, see Fig. \ref{fig:reprodSTATS}). 

Interestingly, SM voxelwise COV values from noise propagation simulations agreed with the measured ones, see Fig. \ref{fig:reprodSTATS}c. This agreement implies that the main factor hindering parameter reproducibility is measurement noise.

\section{Discussion}\label{s:Discussion}
\subsection{Reproducibility}\label{ss:Reproducibility}
\noindent
Biophysical modeling methods for clinically feasibly dMRI protocols must provide reliable parametric maps to achieve adoption in clinic and clinical research. For the first time, we assessed the reproducibility of unconstrained estimation of Standard Model parameters in human white matter. We proposed a framework to optimally design the acquisition protocol by selecting the combination of $b$-values and $\B$-tensors that simultaneously maximize accuracy and precision for SM parameter estimation over a range of biologically plausible values. The optimized protocol, tailored to specific hardware and acquisition time constraints, combined with robust machine learning-based parameter estimation, gives us a favorable position to study reproducibility. 

We report coefficients of variation of SM parameters around $\simeq 5-10\%$ at the voxel-level and around $\simeq 1-4\%$ for ROI means, which are actually comparable to the ones for FA (see Fig. \ref{fig:reprodSTATS}). Noise propagation experiments using the estimated parameters from the WM ROIs showed our predicted COVs from simulation are similar to the ones from experiment ($\pm 50\%$), see Fig. \ref{fig:reprodSTATS}c. Motion and image artifacts contribute to differences with the noise propagation predictions. Nonetheless, approximate COV estimates further support the use of the analytically predicted MSE as a quality metric for optimal protocol design. 

The obtained COVs are encouraging since precise and unconstrained SM estimation has remained elusive for clinically feasible protocols. Unlike previous biophysical modeling works \citep{FIEREMANS2011,ZHANG2012,JELESCU2015b,KADEN2016}
, we remove all parameter constraints. This makes parameter estimation more challenging but reduces biases due to arbitrary constraints and provides additional contrasts that may aid in detecting pathology. Combining optimal acquisition protocols and robust parameter estimation allowed us to obtain SM parameters from 15-minute scans, thereby enabling brain tissue microstructure mapping in clinical settings. 

\subsection{Optimal protocols}
\noindent
The 15-minute optimized acquisition protocols were similar for both scanners albeit with a different TE due to the different maximum gradient strengths for both scanners. After fixing the two low-\textit{b} LTE shells on each protocol ($b=0-1-2 \unit{ms/\mu m^2}$, $\beta =1$) to make them suitable for DTI and DKI analysis, similar complementary shells arose from the optimization for both scanners (see Fig. \ref{fig:optimalProtocols}): A high b LTE shell, an intermediate b - highly anisotropic b-shape shell, and a low b STE shell. It is important to note that the optimal locations for the non-DKI shells depend on the SNR as shown in Fig. \ref{fig:optimalProtocols}b.
Related work from \cite{LAMPINEN2020} optimizing for minimal CRB also revealed the use of non-LTE low \textit{b}-shells to achieve complementary information from DTI/ DKI, \textit{i.e.} $\beta = 0.1 - 0.6$. However, their work also varies TE which limits direct comparison since $T_2$ contrast is also involved.


\subsection{Parameter estimates}
\noindent
The employed machine learning parameter estimation \citep{REISERT2017} is very fast, even for the training step, since the formulation presented by \cite{COELHO2021a} only needs inverting a matrix with averages of the model measurements evaluated over the prior distribution. 
As the ground-truth of all SM-parameters are generally lacking, especially for pathology, we used uninformative prior distributions both for parameter estimation and MSE quantification during the protocol optimization. Such implementation is a significant improvement upon previous works that typically rely on either hard constraints or soft tissue priors \citep{FIEREMANS2011,ZHANG2012,KADEN2016}. 

All SM-parameter values are found to be consistent between subjects (see Fig. \ref{fig:ROIvalues_boxplots}). While ground truth values are unavailable for \textit{in vivo} dMRI experiments, their values are plausible and agree well with prior studies.  The axonal water fraction $f$ showed the largest values in regions with the most densely packed axons while $f_\text{w}$ only has large values close to the ventricles likely due to CSF partial volume. 
The major WM ROIs show $D_\text{a}$ values centered at $\sim 2.3\unit{\mu m^2/ms}$ for the TE=92ms data, in agreement with measurements using more extensive acquisitions involving high diffusion weighted PTE done by \cite{DHITAL2019}, and with high b LTE \citep{VERAART2018}, yet slightly lower than those reported by \citep{HOWARD2020,NILSSON2021}. Furthermore, $D_\text{a}>D_\text{e}^{||}$ in the majority of voxels, agreeing with gadolinium-based contrast experiments in the rat corpus callosum \citep{KUNZ2018}, though this observation depends on the ROI and both values are generally close to each other. 

\subsection{Limitations}
\noindent
As we optimized the measurements for a given scan time, SNR levels may still be in an intermediate regime where some information from the prior ``leaks" into the parameter fits. This may be especially the case for the TE=127ms data and for the noisiest parameters: $D_\text{e}^{||}$ and $f_\text{w}$. 
Of note, both $D_\text{a}$ and $D_\text{e}^{||}$ become smaller for the TE=127ms data, potentially due to the decreased SNR and reduced number of measurements, which results in estimates getting closer to the mean of the training data. In such regime, increased precision comes at the cost of bias. 
Nonetheless, both scan protocols are still able to capture reproducible differences between distinct WM regions for all parameters, Fig. \ref{fig:ROIvalues_boxplots}. Hence, despite the potential introduction of bias for given parameters, the proposed protocols and fitting procedures enable capturing biological variability across WM regions and potentially also variability due to pathology.

Due to duty cycle gradient heating, LTE high-b shells were acquired at $b=5.5\,\unit{ms/\mu m^2}$ rather than $b=8\,\unit{ms/\mu m^2}$ for the Prisma and at $b=5\,\unit{ms/\mu m^2}$ rather than $b=7\,\unit{ms/\mu m^2}$ for the Skyra, as initially suggested by the protocol optimization to avoid increasing the TE (see Fig. \ref{fig:optimalProtocols}a). Noise propagation experiments showed such change did not significantly deteriorate the quality of the protocol. This happens because at high-b the location of shell A is already independent from the rest and the MSE becomes very flat around it, see a similar situation for shell B in Fig. \ref{fig:optimalProtocols}c. 

Overall, the protocol optimization is a high-dimensional problem which has multiple local minima. Our approach showed robust estimations of global optima in a toy function of similar dimensionality and complexity.
We do not explore $\B$-tensor shapes without axial symmetry (inner part of the triangle in Fig. \ref{fig:MDdMRI_TElandscape}a). Releasing this constraint would increase the space of acquisitions but not necessarily increase precision in the parameter estimation, as shown previously for the cumulant expansion parameters \cite{COELHO2019}. Therefore, we constrained $\B$-tensors to be axially symmetric, reducing the dimensionality of the optimization problem. Future work will explore reproducibility of optimal acquisitions with varying TE to simultaneously capture compartmental $T_2$ values, as it was proposed by \cite{VERAART2017,MCKINNON2018} and later assessed by \cite{LAMPINEN2020}.

\section{Conclusion}\label{s:Conclusion}
\noindent
This work provided optimal protocols for WM diffusion modeling and studied the reproducibility of the unconstrained SM in clinical settings. For this, we coupled optimal experimental design with robust parameter estimation and proposed a general framework to obtain the protocol that minimizes the RMSE of the Standard Model parameters. 

\textit{In vivo} experiments performed in twenty normal subjects showed $\lesssim10\%$ COV for all voxel-wise parameter estimates except free water fraction and $\sim 1-4\%$ for ROI means, comparable to DKI derived metrics. These are encouraging results that may boost the application of WM biophysical modeling into clinical research. 

This work reveals that three diffusion measurement settings provide complementary information to DKI: high b LTE, intermediate/high b- highly anisotropic b-shape, and intermediate b STE. Finally, our framework is flexible and can be adapted to different acquisition constraints (\textit{e.g.} scan time, resolution, hardware, etc). All processing codes for the estimation of the Standard Model (SMI toolbox) are available at \href{https://github.com/NYU-DiffusionMRI/SMI}{https://github.com/NYU-DiffusionMRI/SMI}.



\section*{Acknowledgments}
\noindent
This work was performed under the rubric of the Center for Advanced Imaging Innovation and Research (CAI2R, \href{https://www.cai2r.net}{https://www.cai2r.net}), a NIBIB Biomedical Technology Resource Center (NIH P41-EB017183). This work has been supported by NIH under NINDS award R01 NS088040 and NIBIB awards R01 EB027075.

\section*{Conflict of interest}
\noindent
GL, BAA, JV, DSN, EF and NYU school of Medicine are stock holders of MicSi, Inc. - post-processing tools for advanced MRI methods.



\bibliographystyle{elsarticle-harv}
\bibliography{Coelho_bibliography_2020_12_10.bib}

\end{multicols}

\end{document}